# Phonons, Phase Transitions and Thermal Expansion in LiAlO$_2$: An ab-initio Density Functional Study

Baltej Singh[1,2], M. K. Gupta[1], R. Mittal[1,2] and S. L. Chaplot[1,2]

[1]Solid State Physics Division, Bhabha Atomic Research Centre, Mumbai, 400085, India

[2]Homi Bhabha National Institute, Anushaktinagar, Mumbai 400094, India

**Abstract**

We have used *ab*-initio density functional theory technique to understand the phase transitions and structural changes in various high temperature/pressure phases of LiAlO$_2$. The electronic band structure as well as phonon spectra are calculated for various phases as a function of pressure. The phonon entropy used for the calculations of Gibbs free energy is found to play an important role in the phase stability and phase transitions among various phases. A sudden increase in the polyhedral bond lengths (Li/Al-O) signifies the change from the tetrahedral to octahedral geometry at high-pressure phase transitions. The activation energy barrier for the high-pressure phase transitions is calculated. The phonon modes responsible for the phase transition (upon heating) from high pressure phases to ambient pressure phases are identified. Moreover, *ab*-initio lattice dynamics calculations in the framework of quasi-harmonic approximations are used to calculate the anisotropic thermal expansion behavior of γ-LiAlO$_2$.

**PACS numbers:** 63.20.-e, 65.40.–b

**Keywords**: Phonons in crystal lattice, Thermal properties of crystalline solids, High pressure, Phase transition

# I. Introduction:

The application of pressure or temperature on a material may give rise to interesting thermodynamic properties[1-3] and phase transitions[4-7]. The high-pressure experimental techniques have led to the synthesis of many functional materials[8-11] like superconductors, super-hard materials and high-energy-density materials. The pressure introduces compression in the material which gives rise to increased overlap of the electron cloud[7, 12-14]. This leads to a rearrangement of the band structure, which is reflected in the changes in the optical, electrical, dynamical and many other physical properties[15-17]. The change in atomic coordination, symmetry and atomic arrangement[18] takes places in pressure- induced phase transformations[19, 20]. The pressure- volume curve of a material is important in simulating nuclear reactor accidents, designing of inertial confinement fusion schemes and for understanding rock mechanical effects of shock propagation in earth due to nuclear explosion.

$LiMO_2$ (M=B, Al, Ga, In) compounds are widely studied due to their rich phase diagrams. The polymorphism in these compounds arises from the cation order disorder as well as coordination changes[21-23]. The transitions in these systems occur by large lattice mismatch and are accompanied by enormous stresses in the crystals[21, 22]. $LiAlO_2$ finds several important applications. It is used as a coating material for Li based electrodes[24-26] and as an additive in composite electrolytes[27]. This material is a lithium ion conductor[28, 29] at high temperature. A similar compound, $LiCoO_2$, well known for battery cathode applications, is not cost effective[30]. Moreover, $LiCoO_2$ decomposes at high temperature and is problematic due to the toxic nature of cobalt[30]. Therefore, many other materials are being investigated and engineered for the same application. The γ- $LiAlO_2$ is stable upto 1873 K and is cost effective. It is even more stable towards intercalation/deintercalations of Li from the structure[28, 29]. The Li diffusion in this material occurs with a migration barrier of 0.72(5) eV[28, 29]. $LiAlO_2$ exhibits very small lattice changes during lithium diffusion. This property also makes it suitable for use as a substrate material for epitaxial growth of III–V semiconductors like GaN[31]. It is also used as a tritium-breeder material in the blanket for fusion reactors due to its excellent performance under high neutron and electron radiation background[32, 33].

$LiAlO_2$ is highly studied using various experimental and computational techniques[28, 29, 34-39] in recent years for its interesting properties as a Li ion battery material. The stability, safety and performance of battery materials are related to their behavior (thermal expansion and phase transitions) in the temperature and pressure range of interest[40]. The phase diagram of $LiAlO_2$ as obtained from the X- ray diffraction experiments is reported over a range of temperature and pressure[41]. $LiAlO_2$ is known to have six polymorphs of which the structure of only four (α, β, γ and δ) is known[41]. The stable form under ambient conditions is γ-LiAlO$_2$. The α and β polymorphs begin to convert to the γ-polymorph above 700°C[42, 43]. The occurrence of α→ γ LiAlO$_2$ phase transformation is one of the crucial problems for application of this compound as solid phase matrix for electrolyte in molten carbonate fuel cells. The γ-LiAlO$_2$ is strongly considered as a breeder material because of its thermophysical, chemical and mechanical stability at high temperatures. The dilatometer measurements[44] for γ-LiAlO$_2$ yielded values of the anisotropic thermal expansion coefficient to be $\alpha_a=7.1\times10^{-6}$ K$^{-1}$ and $\alpha_c=15\times10^{-6}$ K$^{-1}$, while $\alpha_a=10.8\times10^{-6}$ K$^{-1}$ and $\alpha_c=17.97\times10^{-6}$ K$^{-1}$ are reported from recent neutron diffraction studies[29].

However, the energetic and atomistic picture of the phase transitions is not well understood. Here we report a picture through study of the phase stability and phase transitions among various forms of LiAlO$_2$ as a function of pressure and temperature. Experimentally, isotropic static high pressures are obtained using diamond anvil cell while dynamic pressures are achieved from mechanical or laser shock. Ab- initio density functional calculations provide a very good alternative[8, 11, 16, 45-47] to accurately study the energetics and properties of material under extreme pressure conditions[48]. The ab- initio calculated enthalpy is widely used to obtain the pressure stability region of many crystalline solids[45, 49-51]. To be more accurate, the entropy contribution can be added to enthalpy in order to obtain the free energy of the system[47]. Ab- initio lattice dynamics calculations of phonon spectra[10, 15, 18, 52, 53] can provide the phonon entropy contribution to the free energy of solid. The free energy provides the complete pressure-temperature phase diagram and stability range of crystalline solids[47, 54, 55]. The pressure and temperature dependence of vibrational modes[16, 46, 56, 57, 58, 59] can be used to understand the atomistic picture of phase transition.

We have provided the pressure dependent electronic band structure, phonon band structure and energetic of various phases of LiAlO$_2$. Phonon spectra calculations at various pressures are performed in order to understand the role of phonon entropy in stability of phases and their phase transformation. Pressure dependence of elastic properties is calculated to understand the high-pressure behavior of various phases. Moreover, the mechanism of high temperature phase transitions from high pressure phases to ambient phase is studied in terms of phonon modes instability. The anisotropic thermal expansion is calculated for γ- LiAlO$_2$ in the framework of quasiharmonic approximation.

**II. Computational Details:**

The Vienna based ab-initio simulation package[60, 61] (VASP) was used for structure optimization and total energy calculations. All the calculations are performed using the projected augmented wave (PAW) formalism[62] of the Kohn-Sham density functional theory within generalized gradient approximation[63, 64] (GGA) for exchange correlation following the parameterization by Perdew, Becke and Ernzerhof. The kinetic energy cutoff of 820 eV is adapted for plane wave pseudo-potential. A k-point sampling with a grid of 4×4×4, generated automatically using the Monkhorst-Pack method[65], is used for structure optimizations. The above parameters were found to be sufficient to obtain a total energy convergence of less than 0.1 meV for the fully relaxed (lattice constants & atomic positions) geometries. The total energy is minimized with respect to structural parameters. For lattice dynamics calculations, the Hellman-Feynman forces are calculated by the finite displacement method (displacement 0.03 Å). Total energies and force calculations are performed for the 18, 24, 14 and 14 distinct atomic configurations, resulting from symmetrical displacements of inequivalent atoms along the three Cartesian directions (±x, ±y and ±z), for α, β, γ and δ- phases respectively. The convergence criteria for the total energy and ionic forces were set to $10^{-8}$ eV and $10^{-5}$ eV A$^{-1}$. The phonon energies were extracted from subsequent calculations using the PHONON software[66]. The phonon calculation has been done considering the crystal acoustic sum rule. The phonon spectra in the entire Brillion zone for various phases at different pressures are calculated using the finite displacement lattice dynamical methods. The phonon density of states are obtained by integrating the phonon dispersion curve over 8000 points in the entire Brillion zone. Thermal expansion calculations were performed using the pressure dependence of phonon frequencies[67] in the entire Brillouin zone. The details of the anisotropic thermal expansion calculations are given in section III-H. The phonon density of states for various phases calculated at different pressures are used to calculate the phonon entropy. The only phonon entropy at various temperatures has been included in our calculation of free energy. The Gibbs free energy of a system is calculated using

$$G(T,P) = \Phi(V) + E_{vib} + PV$$

Where $\Phi(V)$ is the static lattice energy, $E_{vib}$ is the phonon energy, and *P, V* and *T* are pressure, volume and temperature respectively.

## III. Results and Discussion

LiAlO$_2$ occursin several phases[41] and the crystal structures are known for the α, β, γ and δ- phase. The γ- phase is the stable phase at ambient condition and crystallizes in a tetragonal structure[41]. It consists of LiO$_4$ and AlO$_4$ tetrahedral units (Fig 1(a)). The α, β and δ- phases are the high pressure- temperature phases[41]. The β-LiAlO$_2$, also called as the low-temperature form[68] which is stable below 0 °C, crystallizes in orthorhombic structure and it also consists of LiO$_4$ and AlO$_4$ tetrahedral units (Fig 1(b)) . In high pressure experiments[41], β- phase can be obtained from the γ- phase at 0.8 GPa and 623K. The stability region for β- phase is quite narrow, so it possibly coexists with other phases[41]. The AlO$_4$ and LiO$_4$ polyhedral units share one edge in γ-phase while these polyhedra are corner shared in β- phase. α- LiAlO$_2$ crystallizes in hexagonal symmetry[42] and consists of LiO$_6$ and AlO$_6$ octahedra (Fig 1(c)). It can be experimentally obtained from γ- phase at high temperature and pressure range (0.5-3.5 GPa and 933-1123 K). α- phase can reversibly transform[42, 43] to γ- phase when heated above 600°C at zero pressure. δ-phase is experimentally obtained, at 4GPa using static pressure[22] and at 9 GPa, by dynamic shock compression[69] of γ- phase. It crystallizes in tetragonal structure and consists of $LiO_6$ and $AlO_6$ octahedral units (Fig 1(d)). The structure of δ- phase is known to pose small amount of Li/Al anti-site disorder. The δ- $LiAlO_2$ is stable up to 773K and it transforms to α and γ- $LiAlO_2$ at higher temperatures. The high pressure reconstructive transition from γ to δ- phase is reported as of displacive nature accompanied by enormous stresses stemming from the lattice mismatch[22]. Table-I gives the experimental and calculated structures of the various phases, which also provides some idea of the structural correlation among the various phases. The calculated structures are found to be in good agreement with the available experimental data.

**A. Electronic Structure**

The electronic band structure calculations (Fig 2) of various phases of $LiAlO_2$ are performed using ab- initio density functional theory method. These calculations are performed with a very high dense k-point mesh of 20×20×20 to ensure the fine curvature of electronic bands. The electronic densities of states are obtained by integrating the band structure over complete Brillion zone. The negative energy of bands signifies the filled valence band while the empty conduction bands have the positive energy. For all the phases, the minimum of conduction band and the maxima of valence band are least separated at zone centre (Γ-point) of the Brillion zone. It can be seen that all the phases show a direct band gap (Fig 2) higher than 4 eV and signify insulating behavior. The γ- phase has the least value of band gap of 4.7 eV among all the phases. Therefore, as compared to other phases, the γ- phase maybe most easily tuned for battery electrode applications, by creating defects in the crystal. As we apply pressure, γ- phase transforms to β- phase with band gap of 4.9 eV. The band gap for δ-phase (5.8 eV) and α- phase (6.2 eV) are further higher than that of γ- and β –phases which may be related to the increasing coordination (4 to 6) of Al and Li atoms in these high-pressure phases. The cathode materials for Li ion battery are required to behave as a good electronic and ionic conductor. Although α- phase has a layered distribution of Li in which Li ionic movement could be favorable, yet the high coordination and high band gap could limit the applications of this phase as a cathode material in the Li ion battery.

The calculations reveal increase in electronic band gap in all the phases of $LiAlO_2$ (Fig 3) on application of hydrostatic pressure. Above 20 GPa, the band gap for γ and β phases shows a sudden increase and takes the value which is equal to that for the δ phase. However, the corresponding phase transitions occur at much lower pressures. The band-gap crossing between α and δ phases occurs at very high pressure above 70 GPa.

DFT usually underestimates the band-gap in insulators and semiconductors even if the exact Kohn-Sham potential corresponding to the exact density in a self-consistent field is used. The GW method[70, 71] (where G is the Green's function and W is the dynamically screened interaction) is more pertinent and provides band-gap of insulators and semiconductors in good agreement with experiment. However, the GW calculations are computationally very expensive. In this paper, our interest is to estimate the change in the band-gap as a function of pressure and between various phases. While the absolute value of the DFT band-gap is systematically in error, we may expect that the change in the band-gap, as a function of pressure and between various phases, would be fairly well reproduced by DFT.

**B. Enthalpy and Phase Transitions**

The total energy and enthalpy for various phases of $LiAlO_2$ are calculated as a function of pressure to find the stability region of these phases at high pressures at zero temperature (while we ignore the zero-point phonon energy). The difference in enthalpy is calculated with respect to that of the ambient pressure stable phase (γ-$LiAlO_2$) is shown in Fig 4. It can be seen that the γ and β phase possess nearly same energy and enthalpy at ambient pressure and hence are likely to be found at ambient pressure conditions. However, with increase in pressure, β phase lowers its enthalpy as compared to that of γ phase and hence is favorably found in this (0.0<P<1.2 GPa) pressure range. As pressure is further increased, enthalpy of α- phase get lowered than that of γ and β- phases. Therefore, the α-phase is favored above 1.3GPa. At further higher pressure, enthalpy shows the lowest value for highest pressure δ-phase. The calculated total energy and enthalpy show several crossovers which may indicate possible phase transitions, among metastable states, namely, from γ- to α, γ- to δ, β- to α, β- to δ and α to δ- phase. The enthalpy crossover shows a phase transition from γ- to α at around 1.3 GPa and γ- to δ phase transition at around 3.2 GPa. The transitions from β- to α and β- to δ are observed from the enthalpy plot at around 1.6 and 3.7 GPa respectively. The critical pressures for the high pressure phase transition as calculated from the enthalpy crossover are only slightly different from the experimental values[41]. However, enthalpy curve implies that β-phase is more stable in comparison to γ- phase at zero pressure (at zero temperature).Experimentally, β- phase was reported to be stable at low temperature[68] (<273K). The α to δ- phase transition is found at about 30 GPa. In order to calculate a more accurate picture of phase stability at finite temperatures, we need to include the entropy and calculate the Gibbs free energy in various phases (Section IIIE). Since these are crystalline solids hence most of the entropy is contributed by phonon vibrations.

## C. Phonon spectra and Free Energy

The phonon dispersion curves along the high symmetry directions in the Brillion zone of various high-pressure phases of LiAlO$_2$ are shown in Fig 6. The various high symmetry directions are chosen according to crystal symmetry[72]. The numbers of phonon branches are different for each phase as these branches are related to the number of atoms and symmetry of the primitive unit cell. The slope of the acoustic phonon modes are related to the elastic properties of the crystal. As seen from Fig 6, the slope of acoustic phonon modes for α, β and δ phases are large as compared to that of γ- phase. This indicates comparatively hard nature of these phases, which is related to their lower volume and high-pressure stability. The highest energy phonon bands signify the polyhedral stretching vibrations of Al-O and Li-O bonds. There exists a significant band gap in the phonon dispersion and phonon density of states (from 80 to 90 meV) for γ- phase which reduces in the first high pressure phase (β- phase) and ultimately disappears in the successive high-pressure phases (α and δ phases). This disappearance of phonon band gap and shifting of highest energy phonon branches towards lower energy signify delocalization of phonon bands due to tetrahedral to octahedral coordination as we go for the high-pressure phases. The lowering of Al-O and Li-O stretching frequencies in the high-pressure phases may come from the longer octahedral bonds as compared to tetrahedral bonds.

The calculated total and partial (atom wise) phonon density of states for the various high pressure/temperature phases of LiAlO$_2$, are given in Fig 5. It is clear (Fig. 5) that different atoms contribute in different energy interval to the total phonon density of states. The comparison of total phonon density of states for various phases reveals a peak like behavior in low energy range for α and δ phases as compared to that of γ and β phases. This will give rise to high entropy contribution to the free energy in α and δ phases. This arises from the low energy peaks in the partial phonon density of states associated with Li atoms. However, the comparison of phonon spectra of γ and β phases shows that the phonon spectrum in the ambient pressure phase (γ- phase) is more populated in the low energy region as compared to that of first high pressure phase (β- phase). This will give rise to higher entropy at ambient pressure for γ- phase than that for β- phase.

The mean squared displacements (MSD) of various atoms, <(u$^2$)> at temperature T are calculated from the partial density of states of various atoms (Fig. 5) using the relation:

$$< u_k^2(T) > = \int \left(n + \frac{1}{2}\right) \frac{\hbar}{m_k E} g_k(E) \, dE$$

Where $n = [\exp\left(\frac{E}{k_B T}\right) - 1]^{-1}$, $g_k(E)$ and $m_k$ are the atomic partial density of states and mass of the $k^{th}$ atom in the unit cell, respectively. The calculated MSD of various atoms as a function of temperature are shown in Fig 7. Li atoms in all the phases have higher MSD values as compared to O and Al atoms due to its lighter mass. Among all the phases, the largest MSD values in α- phase is due to well oriented layered like structure of this phase in which atoms preferably vibrate in a-b plane. There are two Wyckoff sites for O atom in β-phase, both of which have very similar MSD behavior as a function of temperature.

**D. Free Energy and Pressure Temperature Phase Diagram**

The free energy as a function of pressure at different temperatures for the various phases of LiAlO$_2$ is calculated. It can be seen (Fig 8) as temperature increases to 600K, the phonon entropy starts playing a significant role in stabilizing the various phases. Due to this entropy contribution, the free energy of γ- phase is slightly lower than that of β- phase unlike the enthalpy behavior discussed above. Therefore, free energy reflects the stability of γ- phase at ambient pressure conditions. The calculated phase transition pressure for γ to β- phase at 600K is found to be 0.2 GPa. This transition was experimentally obtained at 0.8 GPa and 623K[41]. Therefore, phonon entropy plays an important role in γ to β phase transition. This is a first order phase transition with increase in density from 2.61 to 2.68 g/cm$^3$. However, as the free energies of the γ and β phases are very close, both the phases could coexist at ambient pressure conditions as observed experimentally.

Further, the critical pressure for γ (or β) to α phase transition is obtained to be 0.6 (or 0.7) GPa at 600K. This transition is experimentally reported at 0.5-3.5 GPa and 933-1123 K[41]. The γ to α phase transformation is of first order as it involves volume drop of about 20%. This transition takes place by increasing the coordination numbers of both the Li and Al atoms from four to six (more details in section E). The structure of the α-LiAlO$_2$ is hexagonal containing AlO$_6$ and LiO$_6$ octahedral units. The octahedral units in α phase are found to be square bipyramidal in nature with different planner and axial bond lengths. If the γ to α phase transition is prevented, we find from the energy crossover that γ-LiAlO$_2$ could transform to aδ-LiAlO$_2$ phase around 4GPa and 600K. The δ- phase is experimentally obtained from γ- phase with static pressure 4.0 GPa as well with dynamic shock pressure of 9.0 GPa. The earlier calculations[73] using different pseudo potentials underestimated this value to 2.3 GPa. This phase is also made up of AlO$_6$ and LiO$_6$ octahedra. Free energy crossovers also suggest β to α and β to δ phase transitions as earlier discussed in enthalpy plots. The α to δ- phase transition at very high pressure is found at about 40GPa from the free-energy crossover. The experimental data for this phase transition is not yet available. The complete pressure- temperature phase diagrams, for transition between various phases, as calculated from the Gibbs free energy difference is shown in Fig 9.

### E. Structural Changes During Phase Transition:

We have performed the high- pressure structure optimization for various phases of $LiAlO_2$. These calculations are used to extract the pressure dependence of volume (Fig 10) and bond lengths (Fig 10(b-f)) in all the phases. The different phases of $LiAlO_2$ have different number of atoms in the unit cell, so for the sake of comparison, the volume per atom (Fig 10) is plotted. At around 21 GPa the volume/atom for $\gamma$-$LiAlO_2$, which would be in a metastable state at this pressure, becomes equal to that for $\beta$-$LiAlO_2$. Further at around 25GPa, the volume/atom for both $\gamma$-$LiAlO_2$ and $\beta$-$LiAlO_2$ becomes equal to the highest pressure ($\delta$-$LiAlO_2$) phase. The volume/atoms for $\alpha$-$LiAlO_2$ and $\delta$-$LiAlO_2$ phases become equal at very high pressure of 100GPa. The calculated bond lengths (Fig 10) show that the $AlO_4$ (and $LiO_4$) tetrahedra are not regular in $\gamma$ and $\beta$ phase, due to the different Al-O (and Li-O) bond lengths of the tetrahedral units. The tetrahedral bond lengths decrease with increasing pressure and show a sudden increase at around 25 GPa. This sharp increase in Li-O and Al-O bond lengths signify the transformation of $LiO_4$ and $AlO_4$ tetrahedral units to corresponding $LiO_6$ and $AlO_6$ octahedra of $\delta$-$LiAlO_2$. The two different values of each Li-O (and Al-O) bond lengths implies the irregular nature of the octahedral units, which tend to become regular with further increase in pressure up to 100 GPa. Moreover, the bond-lengths of atomic pairs like Al-Al, Li-Li and Li-Al are well separated in $\gamma$-$LiAlO_2$ and $\beta$-$LiAlO_2$ at ambient pressure. These bond-lengths also show a sudden jump at around 25 GPa and converge to corresponding values for $\delta$-$LiAlO_2$. It is interesting to note that even up to very high pressure of 100 GPa, the bond lengths of $\alpha$-$LiAlO_2$ do not converge to that of corresponding $\delta$-$LiAlO_2$ phase.

### F. Energy Barrier for $\gamma$ to $\beta$ Phase Transition

We have calculated the activation energy barrier for the $\gamma$ to $\beta$ phase transformation by following atomic displacements in the unit cell through the transition path. This approach has been used to study[74-76] continuous phase transitions. We have calculated the activation energy barrier using the nudged elastic band method as implemented in the USPEX software[77]. Group theory analysis using Bilbao crystallographic server[78] shows that the transition from $\gamma$ ($P4_12_12$) to $\beta$ ($Pna2_1$) phase can take place through either of the two common subgroups, namely, $P2_1$ and $P1$, of both the phases. We designate these transition paths as Path-I and Path-II respectively. The group- subgroup transformation from $P4_12_12$ to $P2_1$ occurs through Wyckoff site splitting from 4a (Li), 8b (O) and 4a (Al) in $P4_12_12$ to corresponding all 2a (Li, O & Al) sites in $P2_1$. Similarly, group- subgroup transformation from $Pna2_1$ to $P2_1$ takes place through all 4a (Li, O, Al) sites to all 2a (Li, O, Al) sites. For $\gamma$ to $\beta$-$LiAlO_2$ transition through their common subgroup P1, all the Wyckoff sites in $P4_12_12$ ($\gamma$-$LiAlO_2$) and $Pna2_1$ ($\beta$-$LiAlO_2$) split in 1a sites of corresponding P1 subgroup.

The calculation of minimum energy path for γ to β transition has been performed by transforming the actual structure to their common subgroups P2$_1$ and P1. The transformed lattice parameters (a b c) and fractional atomic coordinates (x y z) are related to the parent phase as given below. The transformation matrix T is obtained using Bilbao crystallographic server[78].

$$(a',b',c')_{P2_1} = (a,b,c)_\gamma T \; ; \quad \begin{pmatrix}x'\\y'\\z'\end{pmatrix}_{P2_1} = T^{-1}\begin{pmatrix}x\\y\\z\end{pmatrix}_\gamma + \begin{pmatrix}0\\3/4\\0\end{pmatrix} \; ; \quad T = \begin{bmatrix}1 & 0 & 0\\0 & 0 & -1\\0 & 1 & 0\end{bmatrix}$$

$$(a',b',c')_{P2_1} = (a,b,c)_\beta T \; ; \quad \begin{pmatrix}x'\\y'\\z'\end{pmatrix}_{P2_1} = T^{-1}\begin{pmatrix}x\\y\\z\end{pmatrix}_\beta + \begin{pmatrix}1/2\\0\\1/2\end{pmatrix} \; ; \quad T = \begin{bmatrix}0 & 0 & -1\\-1 & 0 & 0\\0 & -1 & 0\end{bmatrix}$$

$$(a',b',c')_{P1} = (a,b,c)_\gamma T \; ; \quad \begin{pmatrix}x'\\y'\\z'\end{pmatrix}_{P1} = T^{-1}\begin{pmatrix}x\\y\\z\end{pmatrix}_\gamma + \begin{pmatrix}0\\3/4\\0\end{pmatrix} \; ; \quad T = \begin{bmatrix}1 & 0 & 0\\0 & 0 & -1\\0 & 1 & 0\end{bmatrix}$$

$$(a',b',c')_{P1} = (a,b,c)_\beta T \; ; \quad \begin{pmatrix}x'\\y'\\z'\end{pmatrix}_{P1} = T^{-1}\begin{pmatrix}x\\y\\z\end{pmatrix}_\beta + \begin{pmatrix}-1/4\\0\\1/4\end{pmatrix} \; ; \quad T = \begin{bmatrix}1 & -1 & 1\\0 & 0 & 1\\1 & 0 & 1\end{bmatrix}$$

Going from the initial image in the γ phase (P2$_1$ or P1) to the final image in the β phase (P2$_1$ or P1) involves changes in both the cell parameters and atomic coordinates. A large number of structural images are created between the initial and final configurations. The total energy calculations are performed for these distorted structural images.

The total energies for Path-I and Path-II as calculated using DFT are shown in Fig 11. The abscissa axis in Fig 11 represents various image configurations along the transition pathways. An activation energy per atom of 1.24 eV is required for this transformation through Path-I. Similarly, the activation energy per atom of 1.14 eV is calculated for the Path-II. These calculated values for the γ to β phase transition are lower in comparison to the activation energy barrier of 1.8 eV, for the γ to α phase transition[73], as reported from previous calculations which may be expected since the latter transition also involves large coordination changes.

## G. Phonon Instability and High Temperature Phase Transition

The high pressure phases of LiAlO$_2$ become unstable on heating and transform to the ambient pressure phase[42, 43, 68, 69]. The α- phase, which is used in molten carbonate fuel cells, undergoes α to γ phase transition[42, 43] above 600°C. The slow kinetics of this transition makes it difficult to identify the transition temperature. However, the transformation is generally complete by 900**°C**. In this transition the structural volume expands. We have calculated the phonon dispersion curve (Fig. 12) in the α- phase where we have expanded the unit cell of α- phase to the corresponding volume of γ- phase. Three optic phonon branches are found to become unstable. These phonon modes may be responsible for the high temperature instability of the structure and might lead to α to γ phase transition. The eigenvectors of these modes are analyzed (Fig. 12) at the zone centre of the Brillion zone. These modes are highly contributed by the vibrations of Li atoms with small contribution from the oxygen atoms. The first phonon mode involves the Li vibrations perpendicular to the layers of AlO$_6$ octrahedra and may break the layered structure. The other phonon mode involves the sliding motion of Li in between the successive layers of AlO$_6$ octahedra. These types of motion may cause the diffusion of Li in the 2D layers. This may give rise to the observed low binding energy of Li in α-LiAlO$_2$ as observed experimentally[79]. The diffusion of Li breaks the Li-O octrahedral bonds. In a molten carbonate fuel cell, as the cell works, the temperature increases, this may cause large amplitude vibrations of Li atoms. These vibrations may trigger the phase transition from α to γ phase.

The δ- phase is experimentally known to transform to γ- phase at around 1173K[69]. We have calculated the phonon dispersion for δ- phase (Fig. 13) at expanded volume corresponding to that of γ-phase to understand the mechanism of the phase transition. Two optic phonon branches become unstable which are degenerate at zone centre of the Brillion zone. The eigenvectors of these modes involve (Fig. 13) large amplitudes of oxygen atoms along with some contribution from Al atoms. The Li atoms are found to remain steady in these particular phonon modes. Two of the oxygen atoms out of six in AlO$_6$ octahedra show the outward vibrations which may results in breaking of Al-O bonds. All other four oxygens show bending of Al-O bonds about Al atom. This type of vibrations may give rise to formation of tetrahedral form from the existing AlO$_6$ octahedra. Moreover, the vibration of Al atoms towards the centre of the tetrahedral geometry will give rise to the shortening of Al-O bond lengths. Therefore, both the high-pressure phases transform to the ambient phase through very different mechanisms, one initiated by Li vibrations and the other by O and Al vibrational motion.

**H. Pressure Dependence of Elastic Properties**

We have calculated the pressure dependence of elastic constant tensor (Fig 14) for various phases of LiAlO$_2$. The elastic constants are calculated using the symmetry-general least square method[80] as implemented in VASP5.2. The values are derived from the strain−stress relationships obtained from finite distortions of the equilibrium lattice. For small deformations we remain in the elastic domain of the solid and a quadratic dependence of the total energy with respect to the strain is expected (Hooke's law). The number of components[81] of elastic constant tensor is related to the symmetry of the crystal phase. It is clear from the Fig. 12 that the elastic constants of γ- phase have smaller values at ambient pressure conditions as compared to all other phases. The C$_{33}$ tensile component of γ- phase is largest and increase on applying pressure. This implies that the γ- phase restricts the compression along the tetragonal c-axis. The C$_{11}$ component of the same phase shows softening, although small, with pressure and favors the contraction in the *a-b* plane. On the other hand, the tensile components of the β- phase show hardening along *a* and *b* axis and small softening in *c*- direction. This implies that the *ab* plane of γ- phase behaves similar to *c*-axis of β- phase. The other two high- pressure phases pose large values of elastic constant components due to their higher density and low compressibility. Although α- phase possess higher values of tensile components as compared to that of the δ- phase (highest pressure phase), yet the hardening of these component with pressure is less in α- phase than that in δ-phase. The huge increase in tensile elastic constants of δ-phase with pressure implies high rigidity and low compressibility of this phase. For a crystalline solid, to be stable at any conditions, all the phonon frequencies must be positive and the Born elastic stability criteria must be fulfilled. The Born stability criteria demand the elastic constant matrix to be positive definite which give rise to different simplified expressions for different symmetry structures as follow[81]

Tetragonal (γ & δ Phase) : $C_{11}>|C_{12}|$, $2C^2_{13}<C_{33}(C_{11}+C_{12})$, $C_{44}>0$, $2C^2_{16}<C_{66}(C_{11}-C_{12})$

Orthorhombic (β- Phase): $C_{11}>0$, $C_{11}C_{22}>C^2_{12}$,

$C_{11}C_{22}C_{33}+2C_{12}C_{13}C_{23}-C_{11}C^2_{23}-C_{22}C^2_{13}-C_{33}C^2_{12}>0$, $C_{44}>0$, $C_{55}>0$, $C_{66}>0$

Rhombohedral (α- Phase): $C_{11}>|C_{12}|$, $C_{44}>0$, $C^2_{13}<C_{33}(C_{11}+C_{12})/2$,

$C^2_{14}<C_{44}(C_{11}-C_{12})/2$

For all the phases of LiAlO$_2$, the above stability criteria are found to be satisfied at ambient pressure.

I. **Thermal Expansion Behavior of γ-LiAlO$_2$**

The linear thermal expansion coefficients along the 'a' and 'c' -axes have been calculated within the quasiharmonic approximation[82-84]. The calculations require anisotropic pressure dependence of phonon energies in the entire Brillouin zone[67]. An anisotropic stress of 5 kbar is implemented by changing the lattice constant 'a' and keeping the 'c' parameter constant; and vice versa. These calculations are subsequently used to obtain the mode Grüneisen parameters using the relation[84],

$$\Gamma_l(\nu_{q,i}) = \left(-\frac{\partial \ln \nu_{q,i}}{\partial \ln l}\right)_{T,l'} \quad ; l,l' = a,b,c \ \& \ l \neq l'$$

Where $\nu_{q,i}$ is the frequency of $i^{th}$ phonon mode at wave-vector q in the Brillouin zone. The energy (E) and frequency (ν) of a phonon mode are related by the expression, E= hν (h is Planck constant). Therefore, we may also define,

$$\Gamma_l(E_{q,i}) = \left(-\frac{\partial \ln E_{q,i}}{\partial \ln l}\right)_{T,l'} \quad ; l,l' = a,b,c \ \& \ l \neq l'$$

Where $E_{q,i}$ is the energy of $i^{th}$ phonon mode at wave-vector q in the Brillouin zone. In the tetragonal system, Grüneisen parameters, $\Gamma_a = \Gamma_b$. The calculated mode Grüneisen parameters as a function of phonon energy along different crystal directions are shown in Fig. 15(a).

The anisotropic linear thermal expansion coefficients are given by[84, 85]:

$$\alpha_l(T) = \frac{1}{V_0} \sum_{q,i} C_V(q,i,T) [s_{l1}\Gamma_a + s_{l2}\Gamma_b + s_{l3}\Gamma_c], \quad l = a,b,c$$

Where $s_{ij}$ are elements of elastic compliances matrix, $s=C^{-1}$ at constant temperature T=0 K, $V_0$ is volume at 0K and $C_V(q, i, T)$ is the specific heat at constant volume for $i^{th}$ phonon mode at point **q** in the Brillouin zone. The calculated elastic compliances matrix is given in Table II.

The volume thermal expansion coefficient for tetragonal system is given by:

$$\alpha_V = (2\alpha_a + \alpha_c)$$

The calculated temperature dependence of the lattice parameters is in excellent agreement (Fig. 15(b)) with the recent experimental neutron diffraction data[29]. The calculated anisotropic linear thermal expansion coefficients at 300 K are $\alpha_a$= 10.1× $10^{-6}$ K$^{-1}$ and $\alpha_c$=16.5×$10^{-6}$ K$^{-1}$. They compare very well with available experimental values[29, 44]. The thermal expansion along c-axis is large in comparison to that in the a-b plane. The calculated lattice parameters are in a very good agreement with the available experimental data even at very high temperatures up to 1500 K. This implies that the quasiharmonic approximation for this compound is valid even at very high temperature. The phonon modes around 35 and 55 meV have a large positive value of Grüneisen parameter (Fig. 15(a)). The displacement patterns of two of the zone centre modes in this energy range are shown in Fig 16. The eigenvector of the first mode (36.9 meV) shows (Fig. 16) that the mode is highly contributed by the Li motion in a-b plane along with a small component of $AlO_4$ polyhedral vibrations. So, this mode can give an expansion in the a-b plane. The other mode of 55.7 meV (Fig. 16) has a high value of $\Gamma_c$ and involves Li motion in a-c and b-c planes along with some component of $AlO_4$ polyhedral rotation. This type of mode can give an expansion along the c-axis.

## IV. Conclusions

We have used *ab*-initio density functional theory techniques to calculate structural parameters of various phases of $LiAlO_2$ with variable pressure conditions. The application of pressure leads to structural changes in γ- phase through tetrahedral to octrahedral coordination change about Li/Al atoms, which give rise to the high pressure α and δ phases. The phonon entropy is found to play an important role in phase stability and transitions among various phases. On the basis of calculated free energy, the complete phase diagram of $LiAlO_2$ is obtained. Moreover, the phonon modes which are responsible for phase transition (upon heating) from α to γ are found to be dominated by the Li dynamics. This dynamics at high temperature could give rise to Li diffusion and hence the phase transformation to low density γ phase. On the other hand, phonon mode responsible for δ- to γ- phase transformation, upon heating, is dominated by Al and O atoms dynamics. This is accompanied by breaking of two Al-O bonds of $AlO_6$ octahedra while converting to $AlO_4$ tetrahedra. Moreover, the calculated anisotropic thermal expansion behavior of γ-$LiAlO_2$ *using ab*-initio lattice dynamics agrees very well with experimental measurements and the thermal expansion is highly governed by phonon modes which involve the Li dynamics.

**Acknowledgment**

S. L. Chaplot would like to thank the Department of Atomic Energy, India for the award of Raja Ramanna Fellowship.

TABLE-I: Crystal structure for various phases of LiAlO$_2$ as calculated at zero pressure and temperature. The calculated parameters are comparable with experiments[41].

|  | γ-LiAlO$_2$ | | β-LiAlO$_2$ | | δ-LiAlO$_2$ | | α-LiAlO$_2$ | |
|---|---|---|---|---|---|---|---|---|
| Structure | Tetragonal | | Orthorhombic | | Tetragonal | | Hexagonal | |
| Space Group | P4$_1$2$_1$2 (92) | | Pna2$_1$(33) | | I4$_1$/amd (141) | | R-3m (166) | |
|  | Exp. | Calc. | Exp. | Calc. | Exp. | Calc. | Exp. | Calc. |
| a(Å) | 5.169 | 5.226 | 5.280 | 5.316 | 3.887 | 3.906 | 2.799 | 2.827 |
| b(Å) | 5.169 | 5.226 | 6.300 | 6.335 | 3.887 | 3.906 | 2.799 | 2.827 |
| c(Å) | 6.268 | 6.300 | 4.900 | 4.946 | 8.300 | 8.452 | 14.180 | 14.348 |
| V/Z (Å$^3$) | 41.59 | 43.02 | 40.75 | 41.65 | 31.35 | 32.23 | 32.07 | 33.11 |
| ρ(g/cm$^3$) | 2.615 | | 2.685 | | 3.510 | | 3.401 | |
| α, β, γ (º) | 90, 90, 90 | | 90, 90, 90 | | 90, 90, 90 | | 90, 90, 120 | |
| Z | 4 | | 4 | | 4 | | 3 | |
| Polyhedral units | LiO$_4$, AlO$_4$ | | LiO$_4$, AlO$_4$ | | LiO$_6$, AlO$_6$ | | LiO$_6$, AlO$_6$ | |

TABLE-II: The calculated non-zero components of elastic compliance (GPa$^{-1}$) matrix for the γ- phase at ambient pressure.

| S$_{11}$=S$_{22}$ | S$_{12}$=S$_{21}$ | S$_{13}$=S$_{23}$=S$_{31}$=S$_{32}$ | S$_{33}$ | S$_{44}$=S$_{55}$ | S$_{66}$ |
|---|---|---|---|---|---|
| 0.00984 | -0.00332 | -0.00253 | 0.00785 | 0.01679 | 0.01599 |

FIG 1 (Color online) The crystal structure of (a) γ-Phase, (b) β-Phase, (c) α-Phase and (d) δ- Phase of LiAlO$_2$. The polyhedral units around Li and Al are shown in green and blue color respectively.

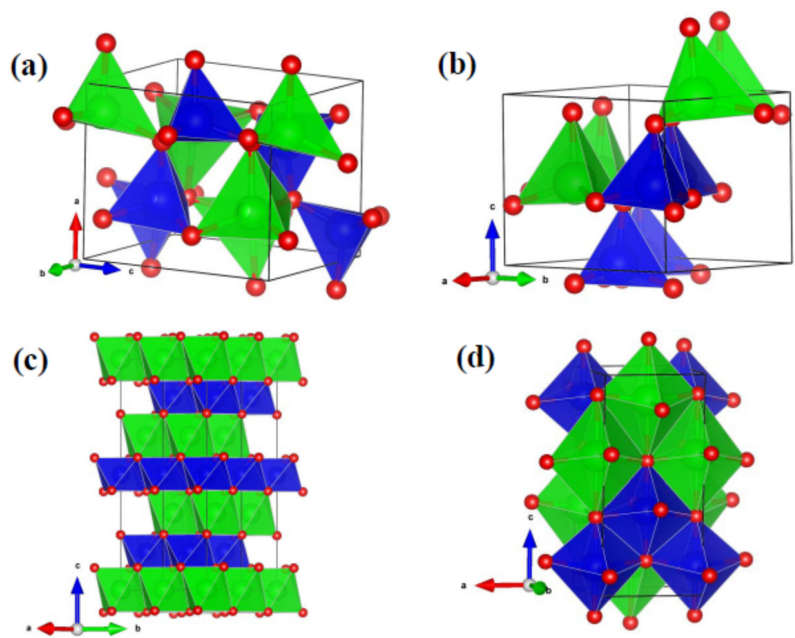

FIG 2 (Color online) Electronic band structure of various high pressure phases of LiAlO$_2$ as calculated using ab- initio DFT. The high-symmetry points are chosen according to the crystal symmetry[72]. EDOS stand for electronic density of states.

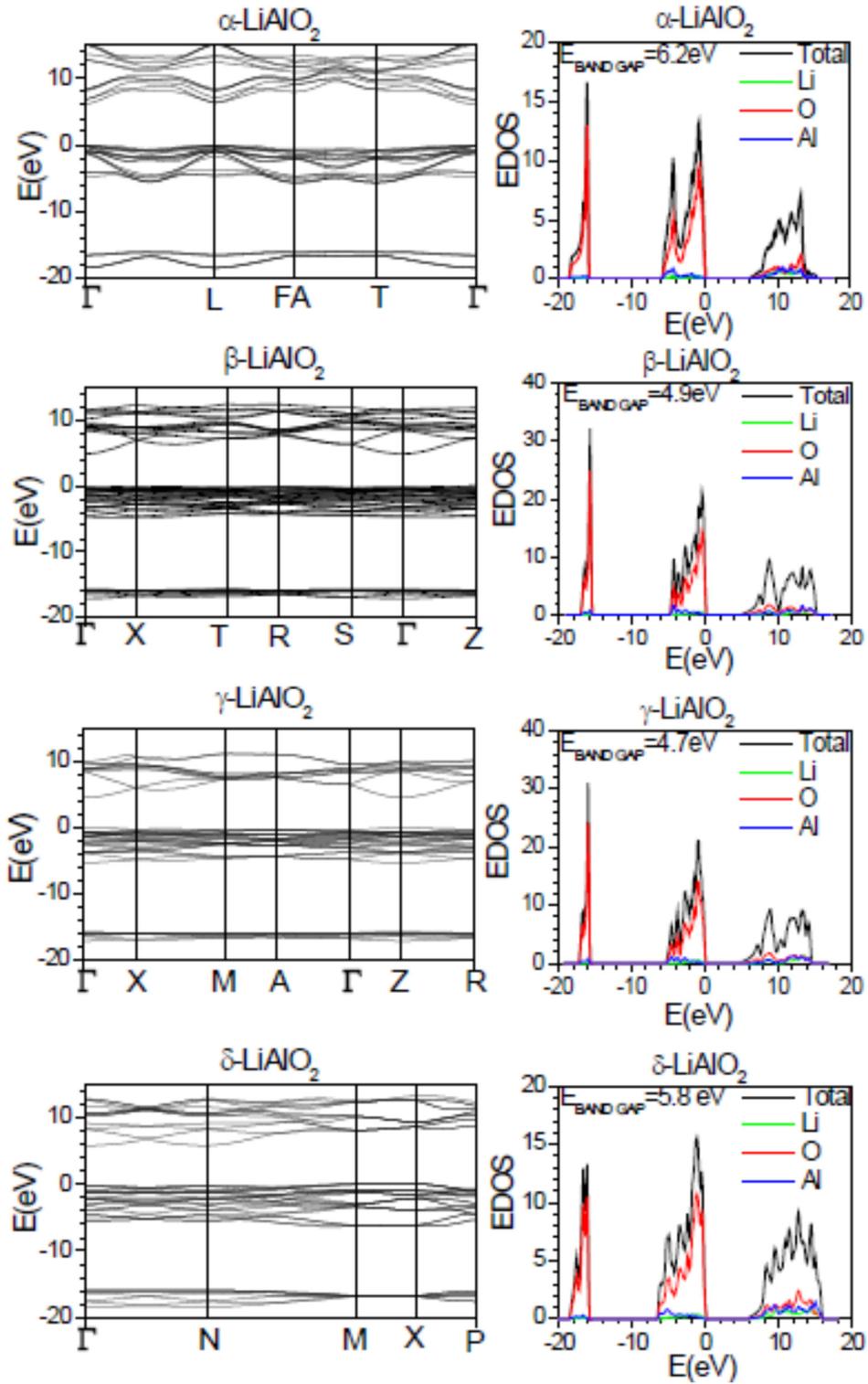

FIG 3 (Color online) Calculated pressure dependent electronic band gap in various phases of LiAlO$_2$ at 0 K.

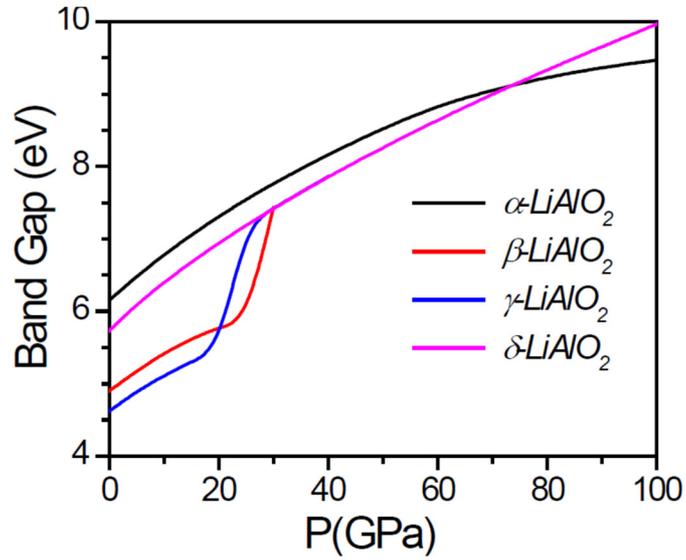

FIG 4 (Color online) Calculated internal energy ($\Phi$), enthalpy (H) and their differences with respect to that of γ-phase, for various phases of LiAlO$_2$, as a function of pressure at 0K.

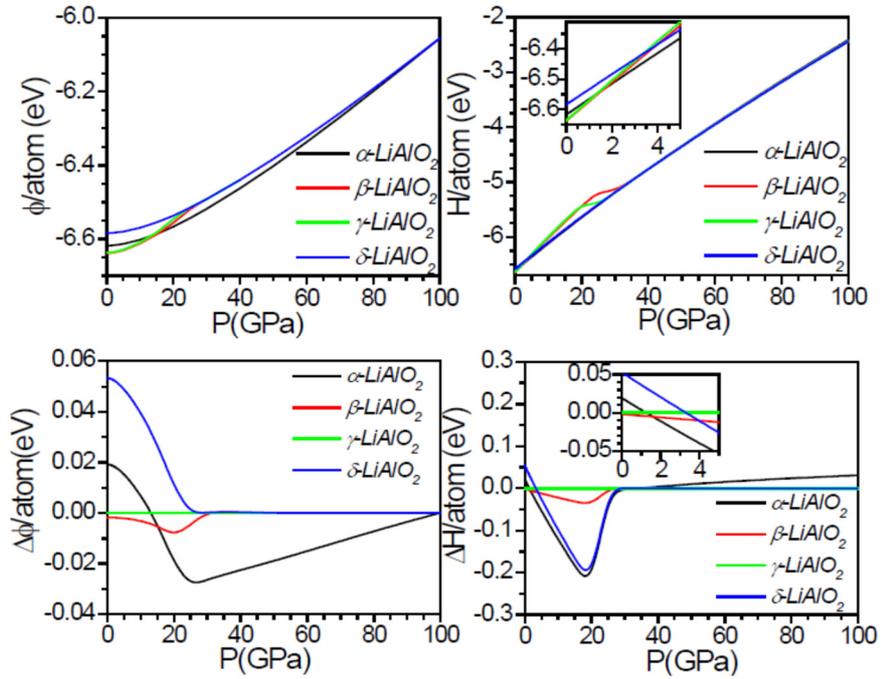

FIG 5 (Color online) Calculated partial and total phonon density of state in various phases of LiAlO$_2$ using ab- initio lattice dynamics.

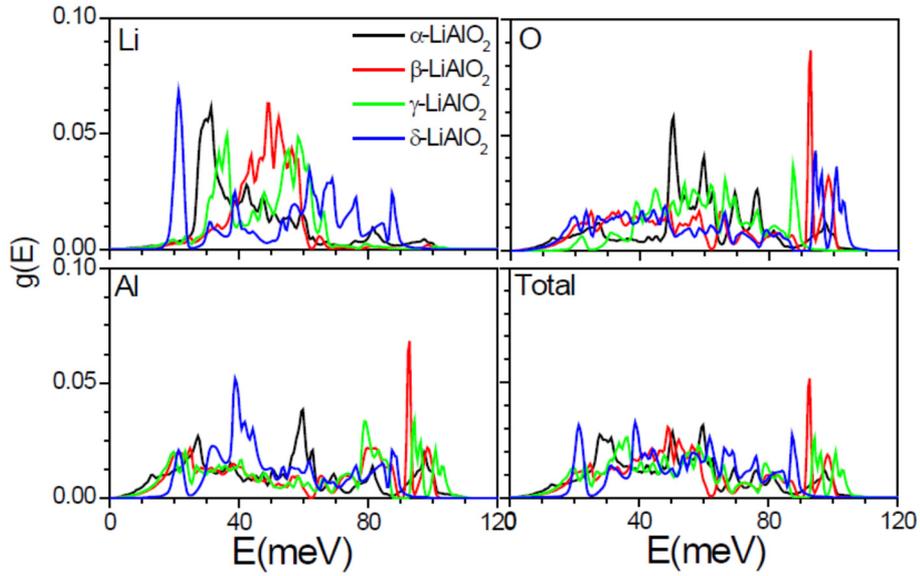

FIG 6 Calculated phonon dispersion curves for various phases of LiAlO$_2$, using ab- initio lattice dynamics. The high-symmetry points are chosen according to the crystal symmetry[72].

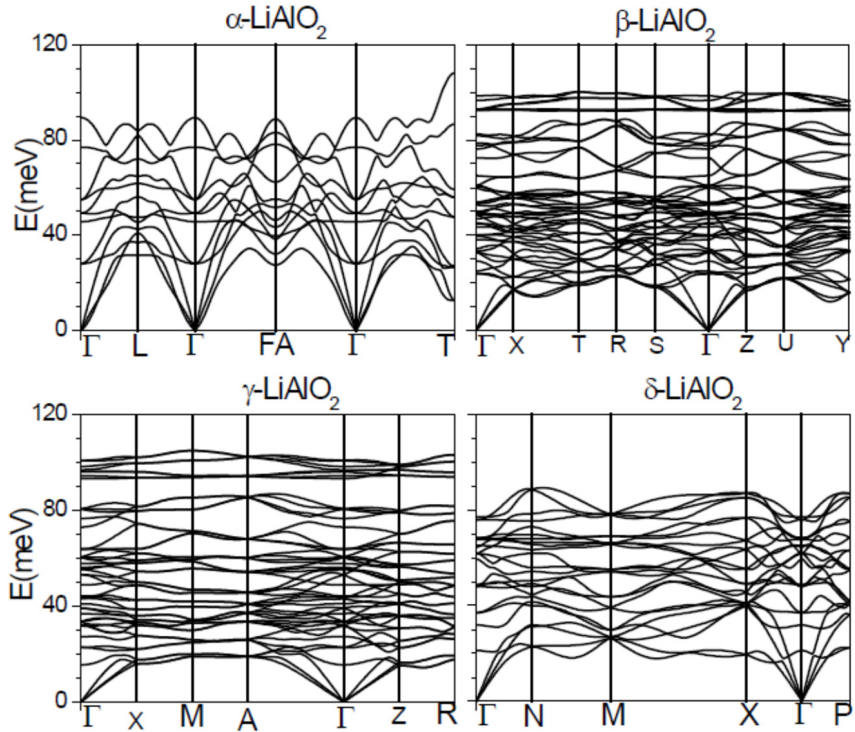

FIG 7 (Color online) Calculated mean-square displacements of atoms in various phases of LiAlO$_2$. The oxygen atoms in the beta-phase occupy two different sites that have slightly different displacements.

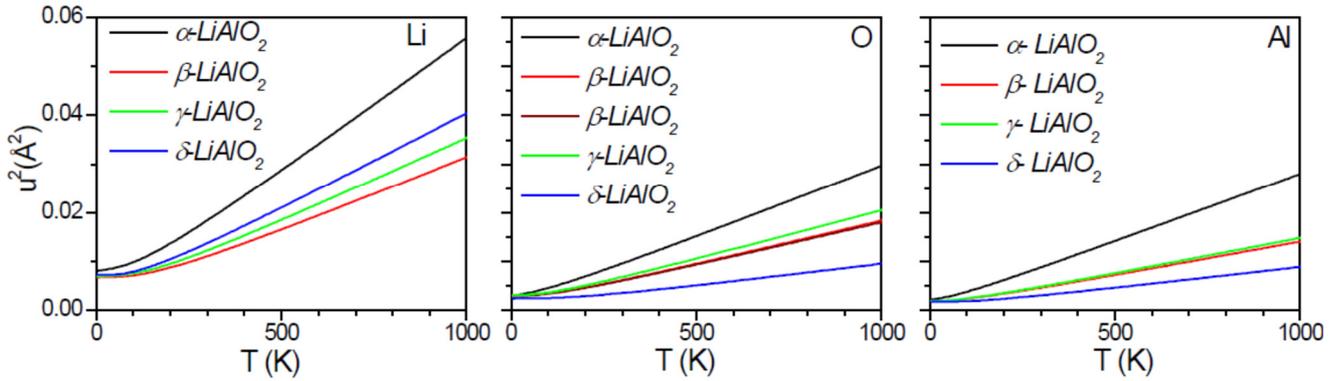

FIG 8 (Color online) Calculated Gibbs free-energy difference for various phases of LiAlO$_2$ compared to the gamma-phase as a function of pressure at 600 K.

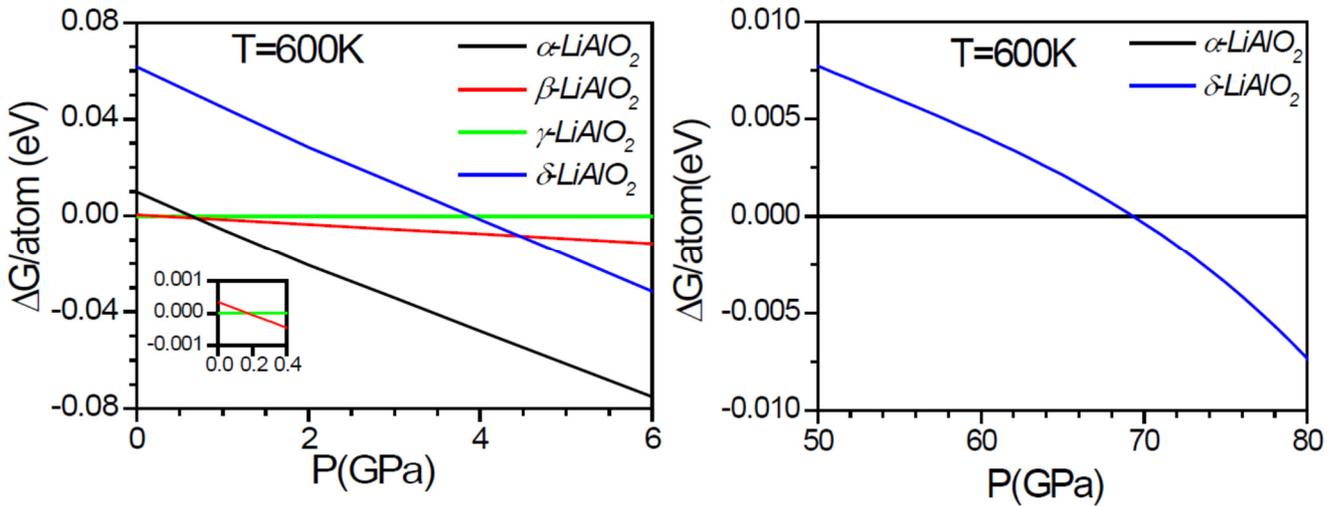

FIG 9 (Color online) Calculated pressure-temperature phase diagram ofLiAlO$_2$. The phase boundaries are calculated from the Gibbs free-energy differences for various phases of LiAlO$_2$.

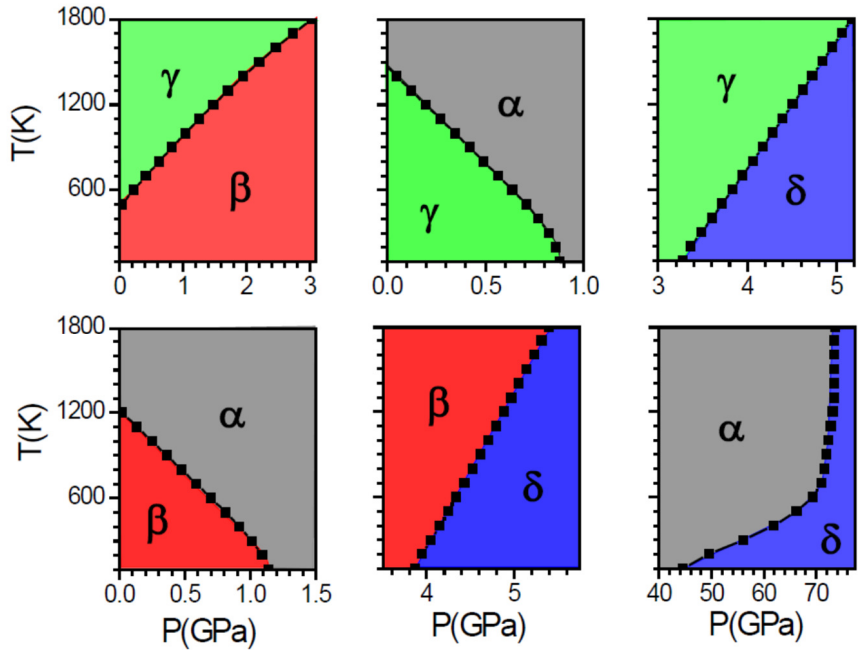

FIG10 (Color online) Calculated pressure dependence of volume and various atomic-pair distances for the various phases of LiAlO$_2$.

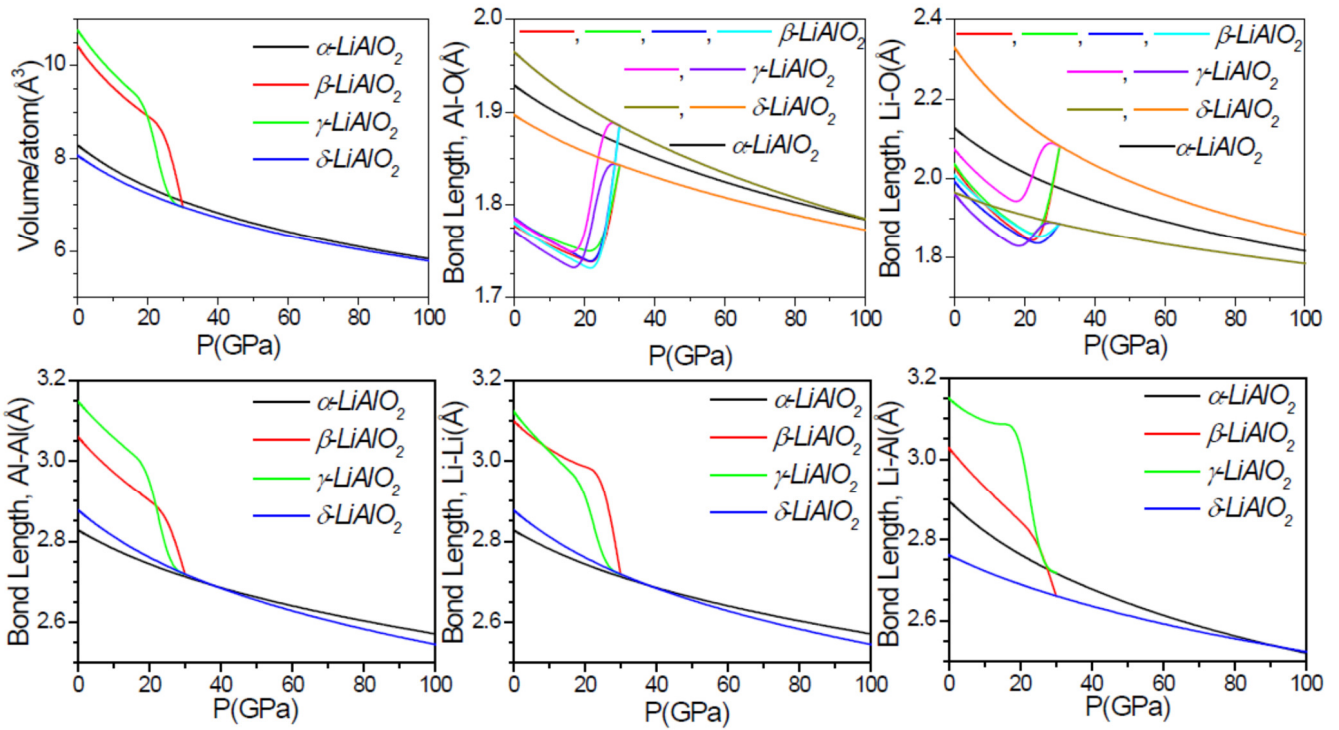

Fig 11. Calculated activation energy barriers through Path-I and Path-II, for γ to β-LiAlO₂ phase transition using ab- initio DFT nudged elastic band method. The abscissa axis in the figure represents various image configurations along the transition pathways through the Path-I and Path-II.

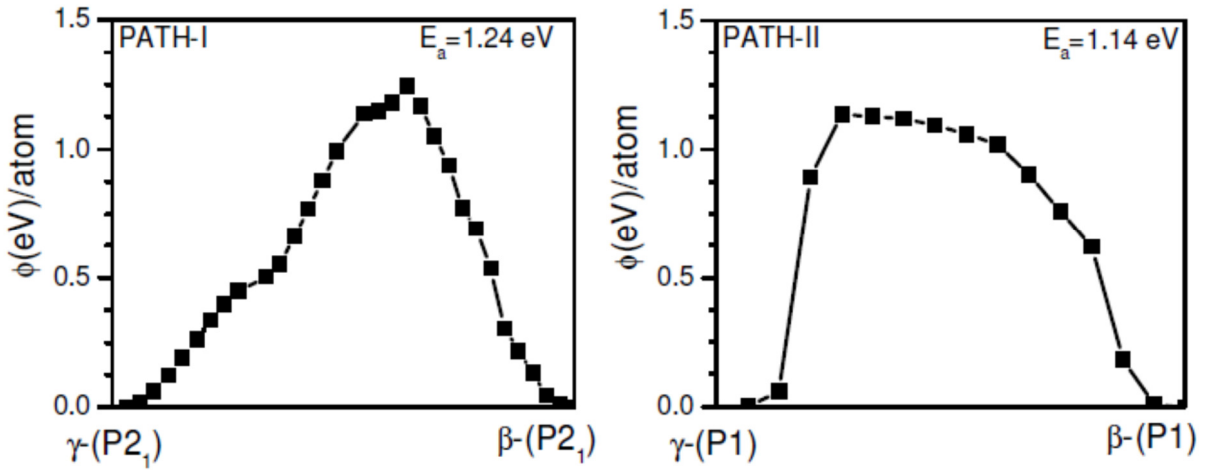

FIG 12(Color online) Calculated phonon dispersion curves of α-LiAlO₂ for the expanded volume. The eigenvector pattern of corresponding unstable phonon modes at Γ point are given on the right. The high-symmetry points are chosen according to the crystal symmetry[72]

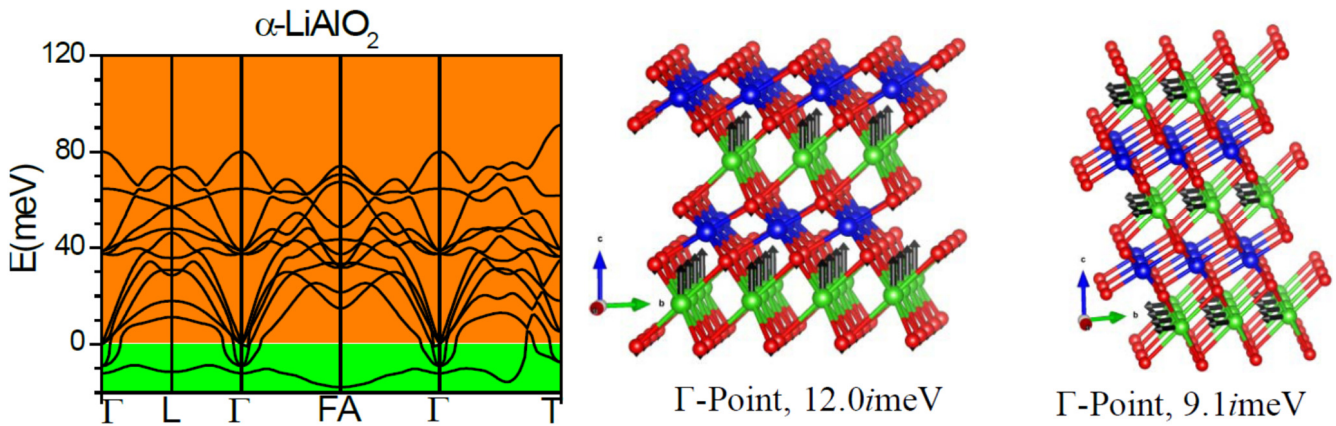

FIG 13 (Color online) Calculated phonon dispersion curves of δ-LiAlO$_2$ for the expanded volume. The eigenvector pattern of corresponding unstable phonon modes at Γ point is given on the right. The high-symmetry points are chosen according to the crystal symmetry[72]

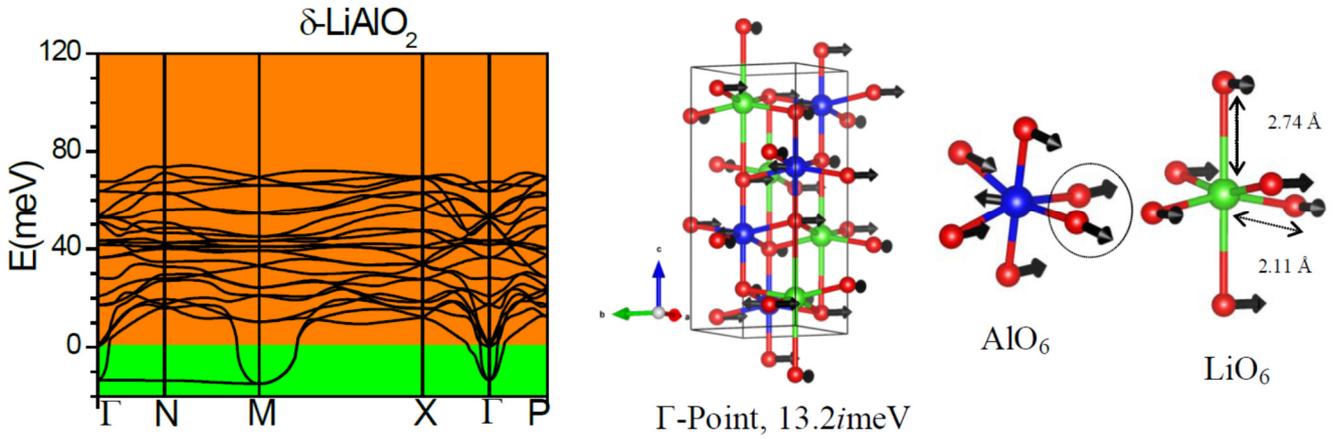

FIG14 (Color online) Calculated pressure dependence of elastic constants in various phases of LiAlO$_2$

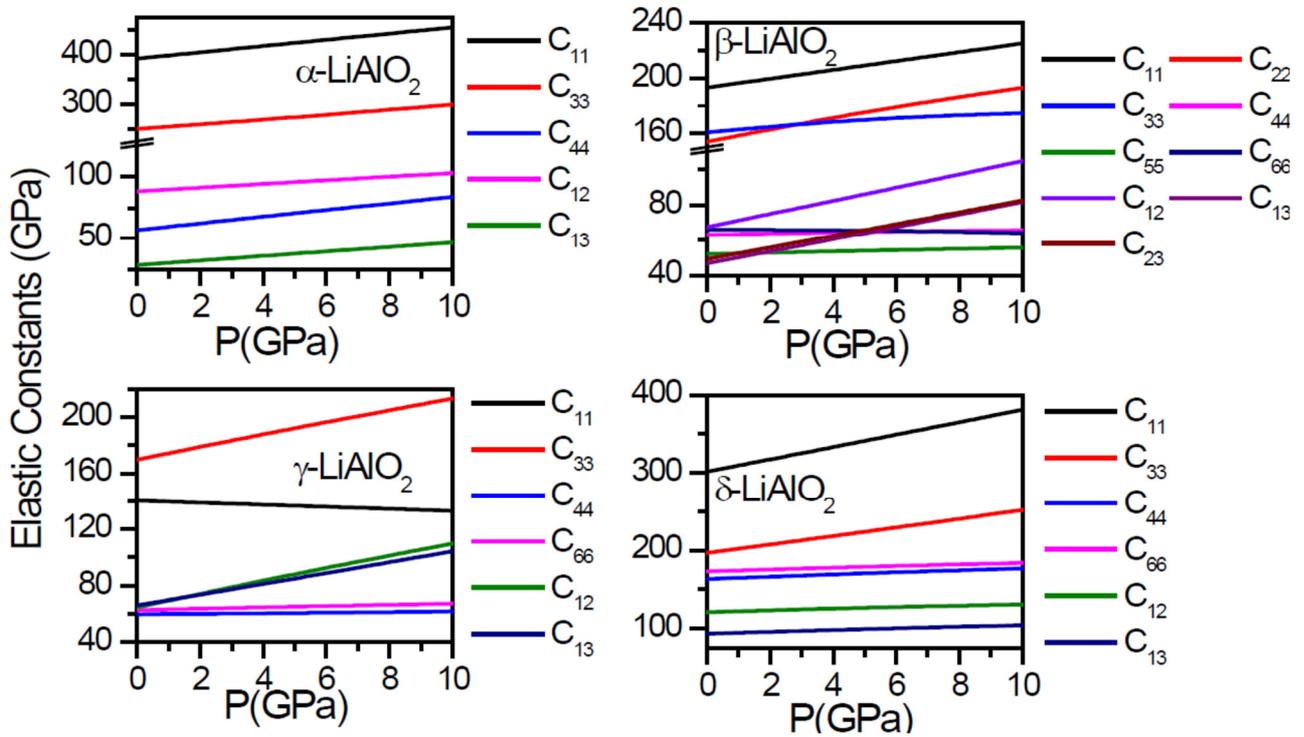

FIG 15 (Color online) (a) The calculated Grüneisen parameter, and (b) the temperature-dependent experimental and calculated lattice parameters and volume [$(l-l_{300 K}/l_{300 K})$, l= a, c, V)].

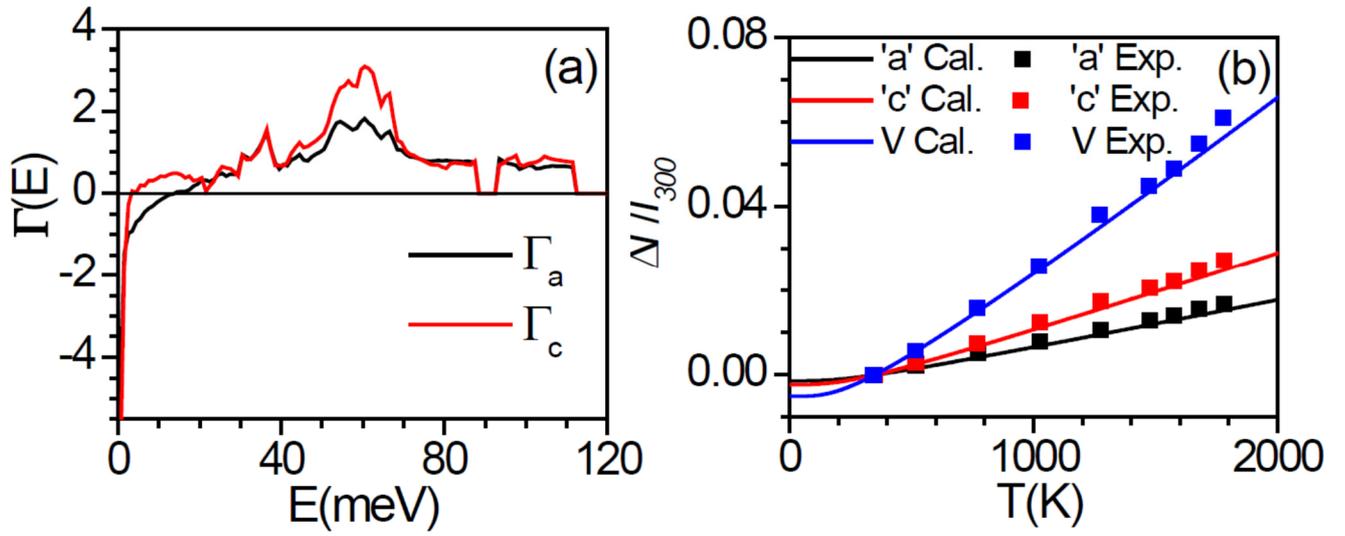

FIG 16 (Color online) The displacements pattern of the zone-centre optic mode which makes a large contribution to thermal expansion in (a) a-b plane, and (b) along c- axis.

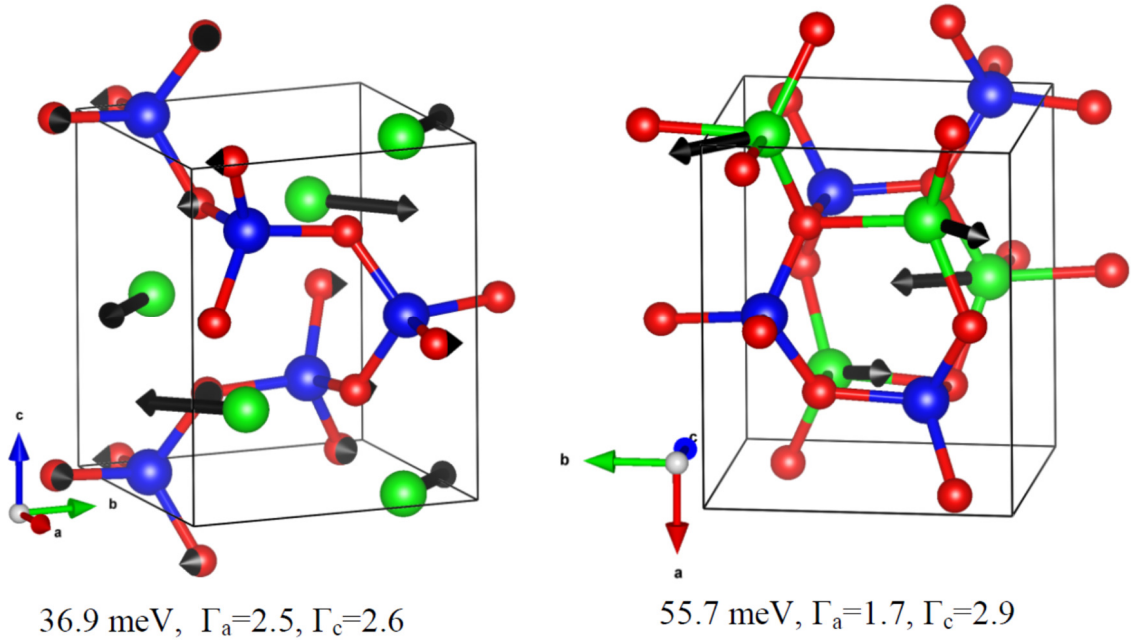

36.9 meV, $\Gamma_a$=2.5, $\Gamma_c$=2.6    55.7 meV, $\Gamma_a$=1.7, $\Gamma_c$=2.9